\documentclass[twocolumn]{revtex4}

\usepackage{framed,xspace}
\usepackage{mathtools}
\usepackage{color}
\usepackage{placeins}
\usepackage{hyperref}

\usepackage{array,url,kantlipsum}

\newcommand{\ave}[1]{\left\langle#1 \right\rangle}

\newcommand{\person}[1]{{\sc#1}}
\newcommand{\concept}[1]{{\sc#1}}

\newcommand{\elabel}[1]{\label{eq:#1}}
\newcommand{\eref}[1]{(Eq.~\ref{eq:#1})}

\newcommand{\Eref}[1]{Equation~(\ref{eq:#1})}

\newcommand{\flabel}[1]{\label{fig:#1}}
\newcommand{\fref}[1]{Fig.~\ref{fig:#1}}

\newcommand{\nmax}{n_\text{max}\xspace}

\newcommand{\ie}{{\it i.e.}~}
\newcommand{\eg}{{\it e.g.}~}

\newcommand{\del}[1]{}

\begin{document}

\title{Evaluating gambles using dynamics}

\author{
Ole Peters$^{1,2}$ and Murray Gell-Mann$^{2}$\\
$^1$London Mathematical Laboratory, 14 Buckingham Street, London WC2N 6DF, UK.\\
$^2$Santa Fe Institute, 1399 Hyde Park Road, Santa Fe, NM, 87501, USA\\
o.peters@lml.org.uk, mgm@santafe.edu}





\begin{abstract}
Gambles are random variables that model possible changes in monetary
wealth. Classic decision theory transforms money into utility through
a utility function and defines the value of a gamble as the
expectation value of utility changes. Utility functions aim to capture
individual psychological characteristics, but their generality limits
predictive power. Expectation value maximizers are defined as rational
in economics, but expectation values are only meaningful in the
presence of ensembles or in systems with ergodic properties, 
whereas decision-makers have no access to ensembles and 
the variables representing wealth in 
the usual growth models do not have the relevant ergodic properties. 
Simultaneously addressing the shortcomings of utility
and those of expectations, we propose to evaluate gambles by averaging
wealth growth over time. No utility function is needed, but a dynamic
must be specified to compute time averages. Linear and logarithmic
``utility functions'' appear as transformations that generate ergodic
observables for purely additive and purely multiplicative dynamics,
respectively. We highlight inconsistencies throughout the development
of decision theory, whose correction clarifies that our perspective is
legitimate. These invalidate a commonly cited argument \cite{Menger1934} for 
bounded utility functions. 
\end{abstract}




\maketitle

\section{Preliminaries}
\label{Preliminaries}
Over the past few years we have explored  a simple but conceptually 
deep and often counter-intuitive change of perspective that 
resolves important problems in economic theory. Here we illustrate this conceptual 
change by resolving the general gamble problem in a very simple setup. 
For clarity of exposition we limit ourselves to the context of an individual 
evaluating gambles in situations where any attendant circumstances other than monetary 
wealth can be disregarded. 

Currently the dominant formalism for treating this problem is utility theory. Utility 
theory was born out of the failure 
of the following behavioral null model: individuals were assumed to optimize 
changes in the expectation values of their wealth. We argue that this null
model is {\it a priori} a bad starting point because the expectation value of wealth 
does not generally reflect what happens over time. We propose 
a different null model of human behavior that eliminates, in many cases, the need 
for utility theory: an individual optimizes what happens to his wealth as time passes. 

The question whether the time average of an observable is well represented
by an appropriate expectation value dates back to the $19\textsuperscript{th}$-century 
development of statistical mechanics \cite{Cohen1997} and is the origin of 
the field called ``ergodic theory.'' In the following we will identify, for different stochastic
processes, stationary independent increments. Being stationary and independent
these observables have many ergodic properties, of which the following specific
property is relevant here.
\vspace{.2cm}

\underline{\bf Ergodic property (equality of averages):}\\ 
The expectation 
value of the observable is a constant (independent of time), 
and the finite-time average of the observable converges to this constant with 
probability one as the averaging time tends to infinity. 
\vspace{.2cm}

Whether an observable possesses this property is crucial when assessing 
the signifcance of its expectation value. We will 
refer to observables with this property as ``ergodic observables.''

Gambles are the formal basis of decision theory. 
Decision theory studies mathematical models of situations that create
an internal conflict and necessitate a decision. For instance we may
wish to model the situation of being offered a lottery ticket. The
conflict is between the unpleasant certainty that we have to pay for
the ticket, and the pleasant possibility that we may win the
jackpot. It necessitates the decision whether to buy a ticket or not. 
Although economics deals with many types of decisions, not all of
which are monetary, the quantitative treatment of the gamble problem
is central to many branches of economics including utility theory,
decision theory, game theory, and asset pricing theory which in turn
informs macroeconomics, as has been argued convincingly
\cite{Cochrane2001}.

We will be dealing with mathematical models but use a common
suggestive nomenclature. In this section we write in \person{small
  capitals} those terms of everyday language that in the following
will refer to mathematical objects and operations. 

A \person{gamble} is a set of possible \person{changes in monetary
  wealth} $\Delta W(n)$ with associated \concept{probabilities}
$p_n(n)$, where $n$ are integers designating \person{outcomes} (or
\person{elementary events}). For convenience, we order
\person{outcomes} such that $\Delta W(n+1)>\Delta W(n)$. Different
\person{gambles} are compared, the decision being which to subject
one's \person{wealth} to, and more generally to what extent.

\person{Gambles} are versatile models, useful to describe a number of
real-world prospects. An insurance contract may be modeled as a
\person{gamble}, as may an investment.  \person{Lotteries} are
important in the historical development of decision theory. Here,
\person{possible payouts} $D(n)$ are purchased for a \person{ticket
  price}, $P$, leading to \person{changes in monetary wealth}, $\Delta
W(n)=D(n)-P$, that are negative up to some value of $n$ and then
positive. This creates a decision problem, when comparing to the
option of \person{doing nothing}: the certain unpleasant prospect of
\person{losing} the \person{ticket price} has to be weighed against
the uncertain prospect of \person{winning} one of the $\nmax$ \person{possible
  payouts}.

\section{Outline}
Section~\ref{The_modern} is a modern treatment of the problem, using
dynamics, that is, we use information about temporal behavior, not
exclusively measure-theoretic probabilistic information. 
Expectation values play a central role in economics, essentially
for two reasons. Firstly, the expectation value of any observable
is, by definition, the average over $N$ instances of the observable, 
in the limit $N\to\infty$. It can therefore be relevant for a member of a 
large resource-sharing group. Decision theory, 
however, is concerned with individuals, not with groups, wherefore we disregard
this first possible reason for using expectation values. Secondly, an observable 
may have the ergodic property mentioned in Section~\ref{Preliminaries}, in 
which case it is informative of what happens to an individual over time. We 
are concerned with the conditions under which this second reason for using 
expectation values is relevant.

The two key
quantities in our treatment are the expected rate of change in wealth,
$\frac{1}{\Delta t}\ave{\Delta W}$, for additive dynamics, and the
expected exponential growth rate of wealth, $\frac{1}{\Delta
  t}\ave{\Delta \ln(W)}$, for multiplicative dynamics. Both quantities
were suggested as criteria to evaluate a gamble, $\frac{1}{\Delta
  t}\ave{\Delta W}$ by Huygens \cite{Huygens1657}, and $\frac{1}{\Delta
  t}\ave{\Delta \ln(W)}$ by Laplace \cite{Laplace1814}, although their dynamic
significance was overlooked, and time scales $\Delta t$ were usually
omitted and implicitly set to 1.

Section~\ref{Historical_decision} discusses the complicated historical
development of these two criteria, which we now briefly summarize. 
It is a partial explanation of the fact that this
perspective has lain hidden from view, despite its otherwise
implausible problem-solving power, and the availability of its
conceptual building blocks for more than 100 years now.
It is necessary to re-tell the history of the problem
because of an important misconception that forbids the modern
perspective. 

Bernoulli \cite{Bernoulli1738} suggested a quantity similar to the
exponential growth rate and called it a ``moral expectation,''
interpreting the logarithm in the exponential growth rate as a
psychological re-weighting that humans apply to monetary amounts. This
presented a very simple criterion -- maximizing the expected
exponential growth rate -- in a very complicated
way. Laplace \cite{Laplace1814} corrected Bernoulli formally, though not
conceptually, writing down {\it exactly} the expected exponential
growth rate, though not pointing out its dynamical significance.
Menger \cite{Menger1934} did decision theory a crucial
disservice by undoing Laplace's correction, adding further
errors, and writing a persuasive but invalid paper on the
subject that concluded incorrectly -- in the language of utility
theory -- that only bounded utility functions are permissible. This
forbade the use of either of the dynamically sensible quantities
because -- forced into the framework of utility theory -- the expected
rate of change in wealth corresponds to a linear (unbounded) utility
function, and the expected exponential growth rate corresponds to a
logarithmic (unbounded) utility function.

That unbounded utility functions are not allowed became an established
result. We ask here {\it why} we cannot use unbounded utility
functions, and find no good reason. The arguments for
the boundedness of utility functions that we found are not
scientifically compelling. A visual representation of the convoluted
history of the problem is shown in \fref{decision_theory}.

We conclude in Section~\ref{Summary_and} that the modern dynamic
perspective is legitimate and powerful. Conceptually, its power derives 
from a new notion of rationality. 
Reasonable models of wealth cannot be
stationary processes. Observables representing wealth then 
do not have the ergodic property of Section \ref{Preliminaries}, 
and therefore rationality must not be defined as maximizing expectation
values of wealth. Rather, we propose as a null model to define rationality as maximizing
the time-average
growth of wealth. This can be done by first converting the non-ergodic processes 
into stationary independent increments per time unit and then 
maximizing expectation values of these (identical to their 
time averages because these observables do have the 
ergodic property of Section~\ref{Preliminaries}). These observables are growth rates, and their
definition is dictated by the dynamics of the wealth model. Because
the gamble problem is phrased without reference to possible causes of
the changes in wealth it is entirely general, wherefore our work
resolves a host of more specific problems in economics, such as the
leverage problem \cite{Peters2011a}, the 300-year-old St Petersburg
paradox \cite{Peters2011b}, and the equity-premium puzzle
\cite{PetersAdamou2013}.
The requirement of boundedness
for utility functions is both unnecessary and detrimental to the formalism of 
decision theory. Our analysis of the history of the problem removes this 
unnecessary obstacle in the way of using physically sensible criteria in 
decision theory. 

\onecolumngrid
\begin{center}
\begin{figure}[h!]
\begin{picture}(300,450)(0,0)
  \put(-100,-20){\includegraphics[width=499.34586pt]{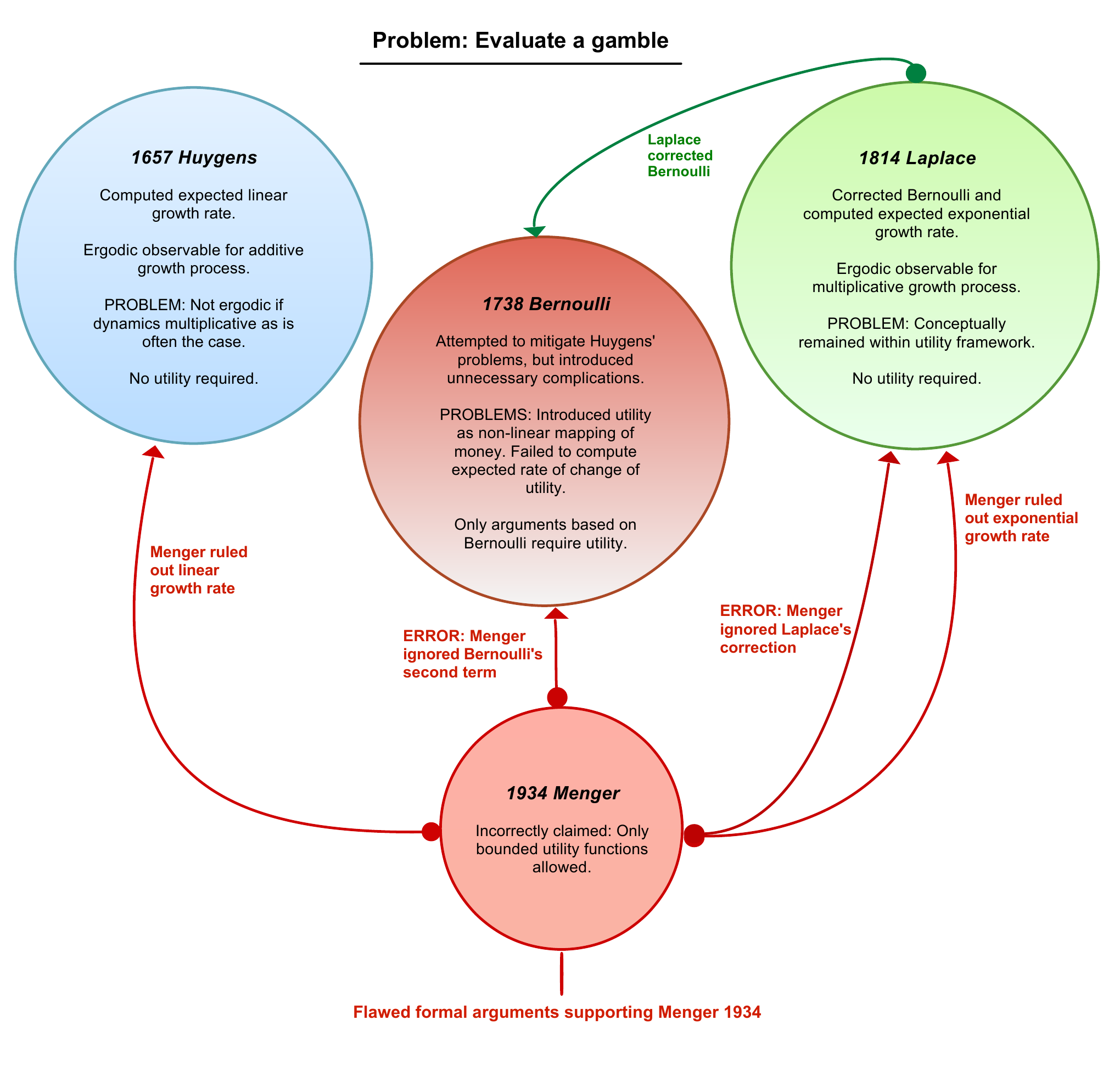}}
\end{picture}
\caption{History of the classic decision theory problem of evaluating
  a gamble. The two physically meaningful solutions are on the left
  and right of the figure. Typically, wealth processes are better
  modeled as multiplicative than as additive, meaning that {\it
    Laplace's Criterion} is usually more relevant, especially when
  changes in wealth $\Delta W$ are of similar scale as wealth $W$
  itself. Problematic aspects are color-coded in red.}
\flabel{decision_theory}

\end{figure}
\end{center}
\FloatBarrier
\twocolumngrid

\section{The dynamic perspective}
\label{The_modern}

In order to evaluate a gamble, we ask: how are the dynamics that the
gamble is part of to be modeled? In the examples below, an
answer to this question allows us to identify
stationary independent increments, \ie to construct 
an ergodic observable, whose expectation value reflects
the behavior over time. Without an answer the problem is
underspecified and cannot be resolved without further assumptions,
for instance about human psychology.

Requesting the specification of a dynamic exposes as underspecified
the original set-up of many problems in economics. Economics treats
randomness in a purely measure-theoretic way: possible outcomes are
given weights (measures, or probabilities), and the overall quality of
a gamble is a weighted average over outcomes, as if all possibilities
were materializing simultaneously with different degrees of
reality. Modern perspectives on randomness actively downplay the
importance of the specific model of measure theory, and emphasize the
need to place the aim of the theory above the conditions imposed by a
specific axiomatization. Thus Gell-Mann and Hartle 
\cite{Gell-MannHartle2012} demonstrate that probabilities beyond 
the interval [0,1] are useful in quantum
mechanics, and Tao \cite{Tao2012} points out the importance of invariance
under extension of the sample space. In our case we argue that a
dynamic is needed in addition to the random variable, turning the
gamble into a stochastic process. Dynamics means repetition, and
requiring the specification of a dynamic is requiring the admission
that we live through time, not in a superverse of parallel worlds with
which we can share resources.

Gambles are often treated in economics as so-called one-shot games,
meaning that they are not part of any dynamic and are assumed to 
reside outside of time, an assumption that is difficult to describe: 
``it's more or less impossible to consider any gamble as happening outside of time'' \cite[p.~3]{Buchanan2013}.
The one-shot setup seems ill-conceived to us, and the methods we propose produce
little insight into the situations it may represent. It is ill-conceived 
because any gamble affects what we may be able to do after the gamble. If 
we lose our house, we cannot bet the house again. The typical decision problem
only makes sense in the context of a notion of irreversible time and dynamics -- 
we cannot go back in time after the gamble, and
our future will be affected by the decisions we make today. One situation that
may be represented by a one-shot game is a bet on a coin toss after which the player 
(who does not believe in an afterlife) will drop dead. Our methods are not developed
for such a-typical cases.

\subsection{Additive repetition}
Treating $\Delta W(n)$ as a (stationary) random variable, repetition
of a gamble may mean different things. Firstly, a gamble may be
repeated additively, so that the wealth after $T$ rounds of the gamble
is
\begin{equation}
W(t+T\Delta t)=W(t)+\sum_{\tau=1}^T \Delta W(n_\tau),
\elabel{additive_repetition}
\end{equation}
where $n_\tau$ is the value of the random variable $n$ in the
$\tau\textsuperscript{th}$ round of the gamble.

$W$  does not have the ergodic property of Section~\ref{Preliminaries}:
the expectation value is not constant in time, 
$\ave{W(t+T \Delta t)}=W(t)+T \ave{\Delta W}$; the finite-time average, 
$\overline{W}_T=\frac{1}{T} \sum_{\tau=1}^{T} W(t+\tau \Delta t)$, does not 
converge to the expectation value but rather
is a random variable whose distribution broadens indefinitely as $T\to\infty$.
Averaging $N$ realizations of $W(t)$ over a large
ensemble ($N\to\infty$) is
not equivalent to averaging $W(t)$ over a long time interval $T \Delta t$ (where $T\to\infty$).

An ergodic observable for \Eref{additive_repetition} exists in
the absolute changes in wealth, $W(t+T\Delta t)-W(t)$, whose
distribution does not depend on $t$. They are stationary independent increments for this dynamic.
The finite-time average of the rate
of change in wealth converges to the expectation value of the rate of change with
probability one,
\begin{equation}
\lim_{T\to\infty}\frac{1}{T\Delta t}\left[W(t+T\Delta t)-W(t)\right]=\frac{1}{\Delta t} \ave{\Delta W(n)}.
\elabel{ergodic_add}
\end{equation}
The expectation value, by definition, is identical to the
ensemble average, $\ave{\Delta
  W(n)}=\lim_{N\to\infty}\frac{1}{N}\sum_{\nu=1}^N
\left[W_\nu(t+\Delta t)-W(t)\right]$, where $W_\nu(t+\Delta t)$
are different parallel realizations of wealth after one round of the
gamble.

This explains {\it Huygens' Criterion}: under additive dynamics as in
\eref{additive_repetition}, the rate of change in wealth is an
ergodic observable, and he who chooses wisely with respect to its
expectation value also chooses wisely with respect to the time
average.

For the specific dynamics of \eref{additive_repetition} an analysis of
this particular observable using our perspective will agree with an
analysis using the economics concept of rationality. This is not the
case for other observables -- {\it Huygens' Criterion} defines something
very special: an {\it ergodic observable} on a {\it non-ergodic wealth
  process}.

Additive dynamics is a simple model, but a moment's reflection reveals
that it is very unrealistic: it assumes that possible changes in
wealth are not affected by the current level of wealth. The
millionnaire and the penniless are modeled as having equal chances of
increasing their respective wealths by \$10,000. For very small gambles ($\Delta
W\ll W$) this linear approximation can be valid (the chances of
gaining \$0.01 may really be similar), and indeed the use
of {\it Huygens' Criterion} emerged from considerations of recreational
gambling where an insignificant amount is bet on a game of dice.

\onecolumngrid
\begin{center}
\begin{figure}[h!]
\begin{picture}(300,375)(85,5)
\put(60,390){(a)}
\put(30,230){\includegraphics[width=184.9500243pt]{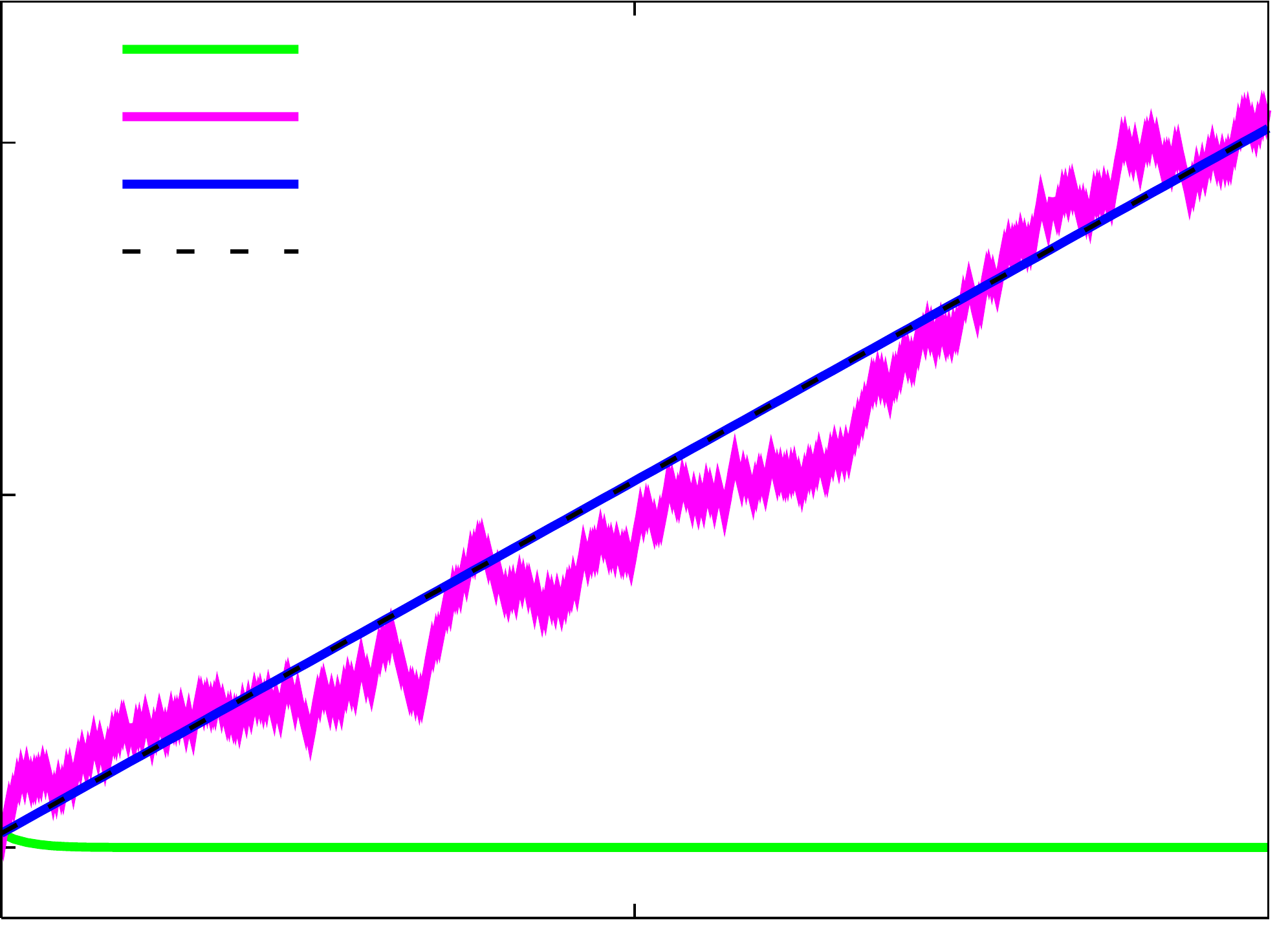}}
\put(-20,290){\includegraphics[width=25pt, angle=90]{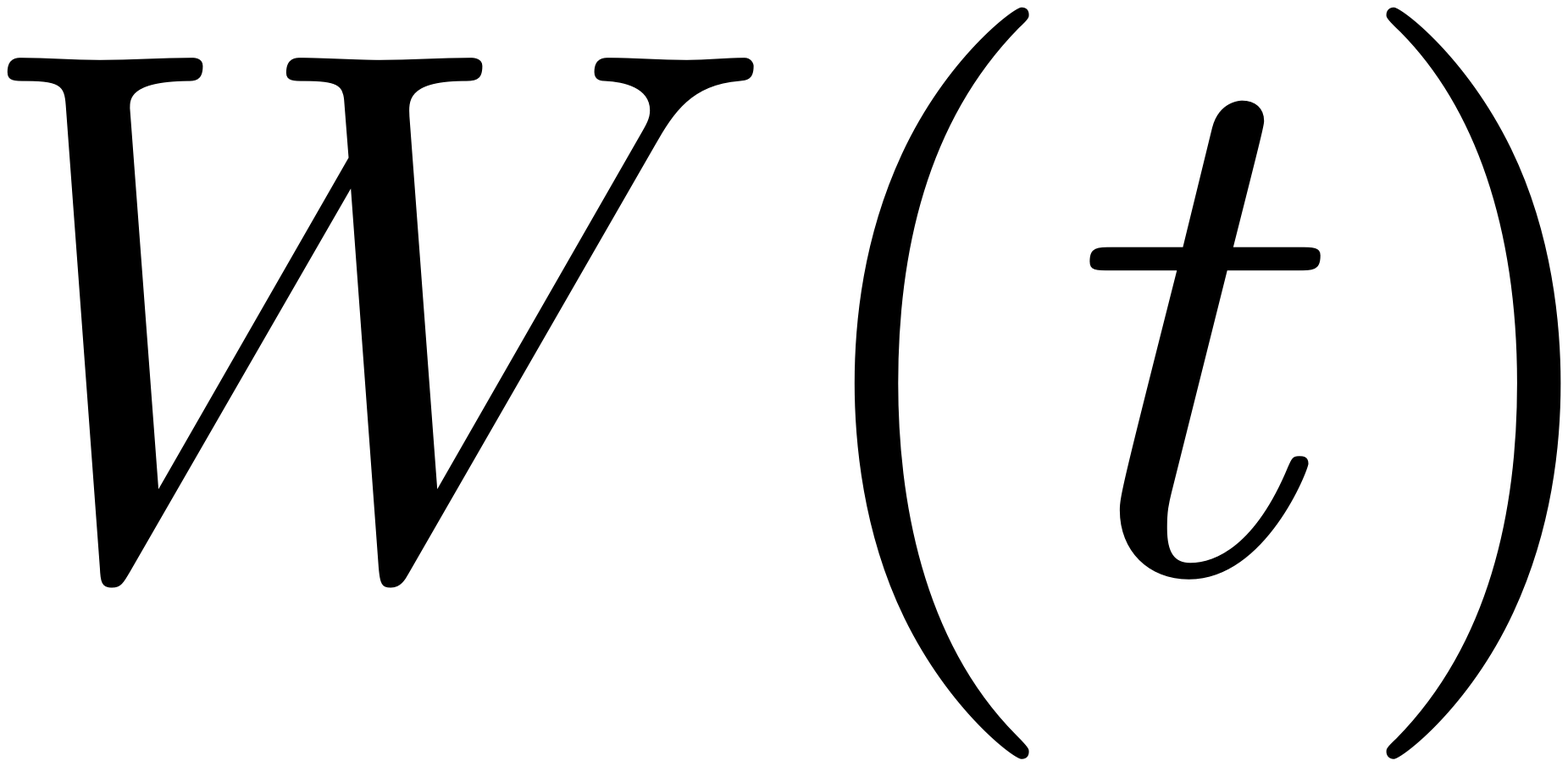}}
\put(80,355){\tiny{$W(t_0)\exp\left((t-t_0)\frac{\ave{\ln(\Delta W)}}{\Delta t}\right)$}}
\put(80,345.2){\tiny{$W(t)$}}
\put(80,335.4){\tiny{$W(t_0)+(t-t_0) \frac{\ave{\Delta W}}{\Delta t}$}}
\put(80,326.6){\tiny{$\ave{W(t)}$}}
\put(15,238){\$0}
\put(10,295){\$25}
\put(10,352){\$50}
\put(30,220){$t_0$}
\put(100,220){$t_0+500 \Delta t$}
\put(190,220){$t_0+1000 \Delta t$}
\put(125,205){$t$}

\put(33.4,210){\line(1,-5){9}}
\put(101,210){\line(2,-1){90}}

\put(42,60){\includegraphics[width=149.803758pt]{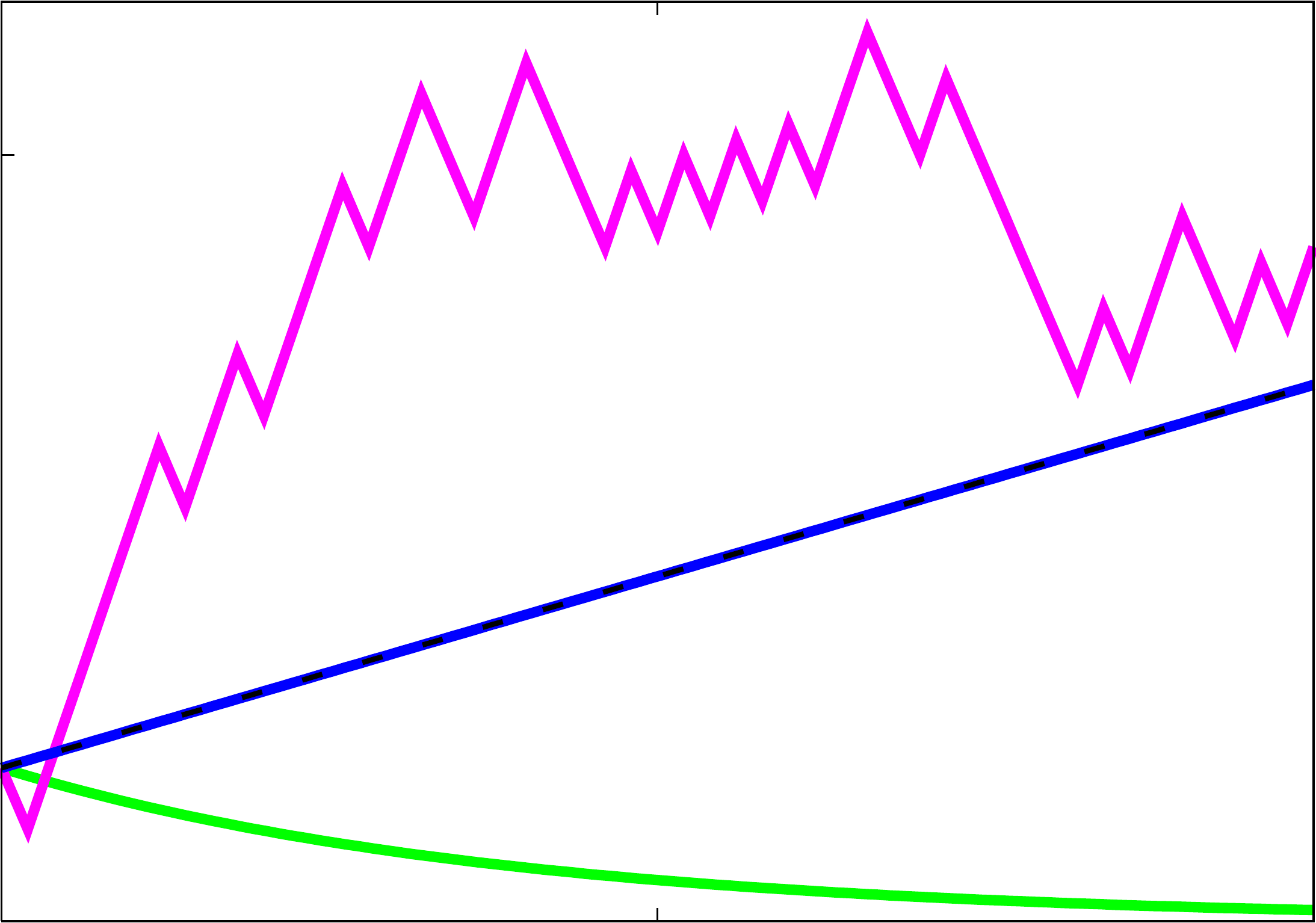}}
\put(-5,105){\includegraphics[width=25pt, angle=90]{ylabel.pdf}}
\put(28,151){\$5}
\put(28,76){\$1}
\put(28,57){\$0}
\put(40,48){$t_0$}
\put(100,48){$t_0+25 \Delta t$}
\put(180,48){$t_0+50 \Delta t$}
\put(123,30){$t$}

\put(320,390){(b)}  
\put(290,230.05){\includegraphics[width=174.11806026pt]{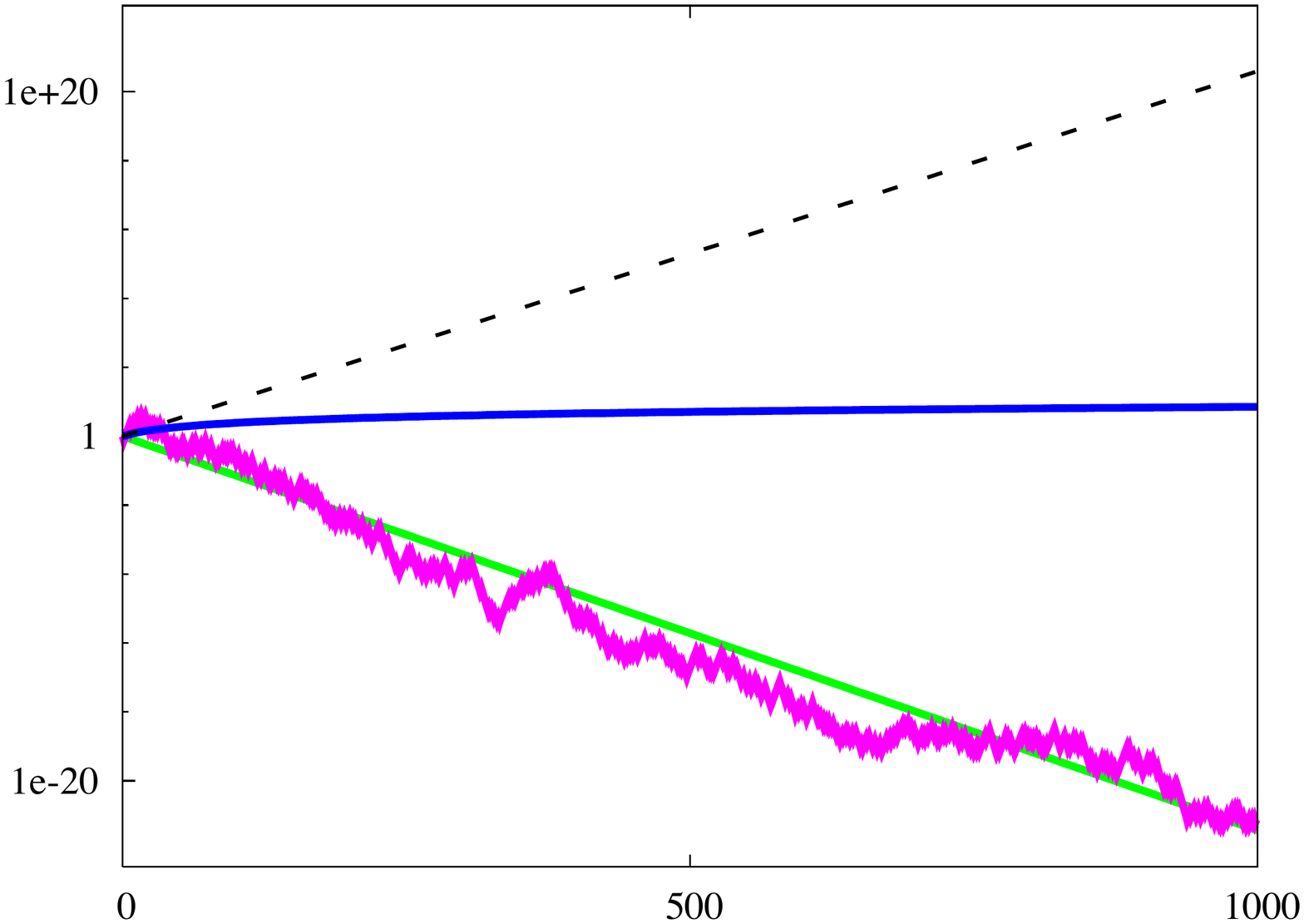}}
\put(275,300){\$1}
\put(260,359){$\$10^{20}$}
\put(255,241){$\$10^{-20}$}
\put(290,220){$t_0$}
\put(355,220){$t_0+500\Delta t$}
\put(432,220){$t_0+1000\Delta t$}
\put(380,205){$t$}

\put(293,210){\line(1,-5){9}}
\put(355,210){\line(2,-1){90}}

\put(302,60){\includegraphics[width=143.6579628pt]{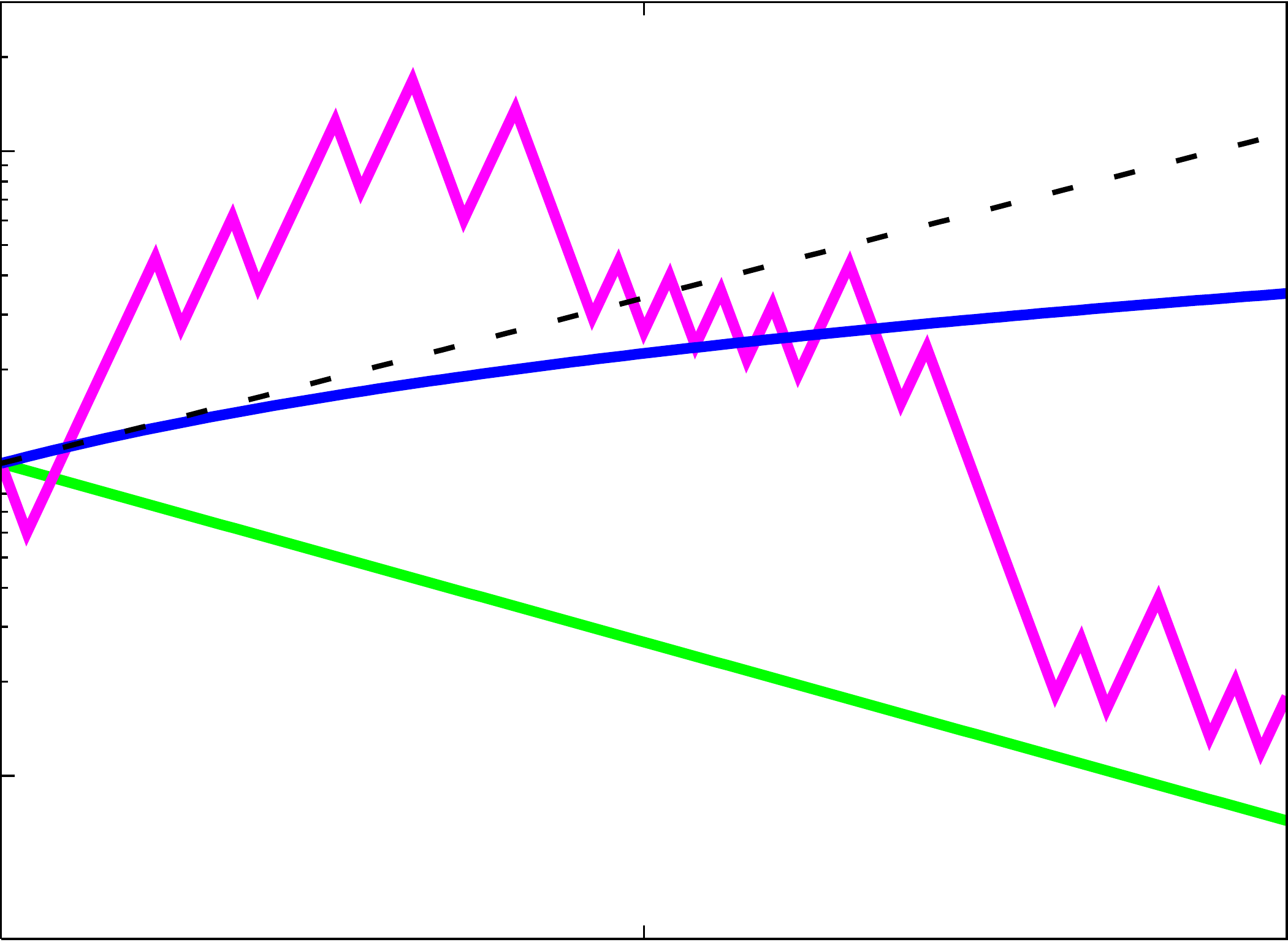}}
\put(279,146){$\$10$}
\put(285,110){$\$1$}
\put(276.1,74){$\$0.1$}
\put(300,48){$t_0$}
\put(350,48){$t_0+25 \Delta t$}
\put(430,48){$t_0+50 \Delta t$}
\put(380,30){$t$}

\end{picture}
\caption{
Assume initial wealth $W(t_0)=\$1$ and toss a fair coin. If
  tails shows ($n=1$), $W$ decreases to $W(t_0+\Delta t)=\$0.60$. If
  heads shows ($n=2$), $W$ increases to $W(t_0+\Delta t)=\$1.50$. The
  gamble is repeated (a) additively, linear plot, and (b)
  multiplicatively, log-linear plot (zoom-ins below the main
  panels). For clarity, the same sequence of heads and tails is
  used in both plots, and the color-codings are identical. A typical
  trajectory is shown (magenta lines). Under (a) the expectation
  value of $W$ (dashed line) grows in time with the expected rate of
  change (ergodic observable for this dynamic, blue line), and a
  trajectory growing exponentially at the expected exponential growth
  rate (green line) does not describe the long-time behavior. Under
  (b) the expectation value of $W$ grows exponentially but has nothing
  to do with the long-time behavior -- $W$ typically decays
  exponentially in this case, following the expectation value of the
  exponential growth rate (ergodic observable for this
  dynamic). The probability distribution of $W$ is concentrated near
  the green line at any $t$, while a small number of a-typical trajectories 
  let the mean grow along the dashed line.
  Linear growth in time at the expected rate of change in
  $W$ does not describe the long-time behavior.} \flabel{coin_toss}
\end{figure}
\end{center}

\FloatBarrier
\twocolumngrid

\subsection{Multiplicative repetition}
A gamble may also be repeated multiplicatively. To simplify notation,
we define per-round growth factors $r(n)=\frac{W(t_0)+\Delta
  W(n)}{W(t_0)}$, where $t_0$ is the time just before the first round
of the gamble. These inherit their stationarity and independence from $\Delta W(n)$. In this case,
\begin{equation}
W(t+T\Delta t)=W(t) \prod_{\tau=1}^T r(n_\tau),
\elabel{multiplicative_repetition}
\end{equation}
which may be re-written as
\begin{equation}
W(t+T\Delta t)=W(t) \exp\left[\sum_{\tau=1}^T  \ln\left(r(n_\tau)\right)\right].
\end{equation}

$W$ again lacks the ergodic property of Section~\ref{Preliminaries}: 
the expectation value is not constant in time, $\ave{W(t+T \Delta t)}=W(t)\exp[T \ln(\ave{r})]$;
the finite-time average 
$\overline{W}_T=\frac{1}{T} \sum_{\tau=1}^{T} W(t+\tau \Delta t)$
does not converge to the expectation value but rather
is a random variable whose distribution is not stationary.
Again, averaging many realizations of $W(t)$ over a large
ensemble ($N\to\infty$) is not equivalent to averaging $W(t)$ over a long time 
interval ($T\to\infty$). Crucially, and in contrast to
the additive dynamic \eref{additive_repetition}, absolute changes in wealth are
not ergodic either. The construction of an ergodic observable now
requires a non-linear transformation.

An ergodic observable for \eref{multiplicative_repetition} exists in
the {\it relative} changes in wealth, $\frac{W(t+T\Delta t)}{W(t)}$,
whose distribution does not depend on $t$.  
Increments in the logarithm of $W$ are now stationary and independent.
The finite-time average of the rate of change in the logarithm of wealth, \ie the
exponential growth rate, converges to the expectation value 
of the rate of change in the logarithm with
probability one,
\begin{equation}
\lim_{T\to\infty}\frac{1}{T\Delta t}\ln \left(\frac{W(t+T\Delta t)}{W(t)}\right)=\frac{1}{\Delta t} \ave{\Delta \ln W(n)}.
\elabel{ergodic_mult}
\end{equation}
The expectation value, by definition, is identical to the
ensemble average, $\ave{\Delta \ln(W(n))}=\lim_{N\to\infty}\frac{1}{N}
\sum_{\nu=1}^N\ln \left(\frac{W_\nu(t+\Delta
  t)}{W(t)}\right)$. 

This explains {\it Laplace's Criterion}: under multiplicative
dynamics, the rate of change in the logarithm of wealth is an ergodic
observable, and he who chooses wisely with respect to its expectation
value also chooses wisely with respect to the time average.
Multiplicative repetition is exemplified by geometric Brownian motion,
the most influential model in mathematical finance.

Multiplicative repetition usually resembles real wealth processes more closely 
than additive repetition. For instance, under multiplicative repetition the likelihood of
a \$10,000 increase in wealth is no longer independent of initial wealth. Instead of 
modelling the millionnaire and the penniless as having equal chances of gaining \$10,000,
multiplicative repetition models them as having equal chances of increasing 
their respective wealths by 1\%. Multiplicative repetition treats zero wealth as a special state, 
resembling a no-borrowing constraint: betting more than we have can have
qualitatively different consequences from betting less than we have.
Another important, subtle, and more realistic property of 
multiplicative repetition is this: time-average growth is impaired by fluctuations. 
Unlike under additive repetition, the introduction of fluctuations that do not affect the 
expectation value does reduce the growth rate that is observed with probability one 
over a long time. In the sense that some dynamical models are more realistic than others, 
reality imposes a dynamic and corresponding ergodic growth rates on the decision-maker.
We may choose which gamble to play but we do not get to choose the mode of repetition.

Equipped with these modern tools, we have no need for the concept of
utility. The general gamble problem can be resolved without it, as can
-- of course -- special cases, such as the 300-year-old St Petersburg
paradox \cite{Peters2011a}.
\vfill\eject

\begin{framed}{\it \bf \underline{Common error}}\\
{\it \textcolor{red}{ Prominent texts in decision theory make
   incorrect statements about the equality of
    expectation values and time averages, as for instance in the
    following passage: ``If a game is `favorable' from the point of
    view of the expectation value and you have the choice of repeating
    it many times, then it is wise to do so. For eventually, your
    amount of money and, consequently, your utility are bound to
    increase (assuming that utility increases if money increases),''
    \cite[p.~98]{ChernoffMoses1959}.}}
\end{framed}

This statement by Chernoff and Moses is not true if ``favorability'' is
judged by an observable that does not have the ergodic property of 
Section~\ref{Preliminaries}.
The
general falsity of their statement is evident in panel (b) of
\fref{coin_toss}, an example of the multiplicative binomial process,
studied in detail by Redner \cite{Redner1990}. Here, $W$ is not ergodic, and the
game is ``favorable from the point of view of the expectation value''
of $W$, but it is certainly {\it not wise} to repeat it many times. We
will use red text in square boxes to highlight errors and weaknesses
in arguments that are commonly believed to be valid.

\section{Historical development of decision theory}
\label{Historical_decision}
In this section we relate the modern treatment of the gamble problem
to classic treatments and highlight common misconceptions as well 
as inconsistencies within the classic view. Aspects 
of this history, which are included here for completeness, are also discussed 
in \cite{Peters2011b}. Menger's ill-conceived rejection
of unbounded utility functions was discussed in \cite{Peters2011c}.
Its treatment here is briefer and more accessible to those unfamiliar 
with his original text.

\subsection{Pre-1713 decision theory -- expected wealth}
Following the first formal treatment by Fermat and Pascal 
\cite{FermatPascal1654} of
random events, it was widely believed that gambles are to be evaluated
according to the expected rate of change in monetary wealth. To give
it a label, this criterion may be attributed to Huygens \cite{Huygens1657},
who wrote ``if any one should put 3 shillings in one hand without
telling me which, and 7 in the other, and give me choice of either of
them; I say, it is the same thing as if he should give me 5
shillings...''

\underline{\it \bf Huygens' Criterion:} \\ {\it Maximize the rate of
  change in the expectation value of wealth,
\begin{equation}  
\frac{1}{\Delta t}\ave{\Delta W(n)}.
\elabel{Huygens}
\end{equation}}
In modern terms, Huygens suggested to maximize the ergodic growth rate
assuming additive dynamics.

Nicolas Bernoulli, in a letter to Montmort \cite{Montmort1713} challenged this
notion by introducing a lottery whose expected payout,
$\ave{D(n)}$, diverges positively with the number of possible outcomes, 
$n_{\text{max}}$. Since the expected rate of change
in wealth $\frac{1}{\Delta t}\ave{\Delta W(n)}=\frac{1}{\Delta
  t}\left(\ave{D(n)}-P\right)$ is linear in $\ave{D(n)}$, it too
diverges for any finite ticket price $P$.  According to {\it Huygens'
  Criterion} any finite ticket price should be paid for the
lottery. However, N. Bernoulli chose the lottery such that
large gains only occur with small probability, and found that typical
individuals when (hypothetically) offered this lottery were
not willing to pay much. This seeming incongruence became
known as the St Petersburg paradox. 
From the modern perspective, the paradox challenges the notion of
rationality defined as expectation-optimization, or the assumption of
unrealistic additive dynamics.
It exposes

\begin{framed}
\underline{\it \bf Huygens' weakness}\\ {\it \color{red} Expectation
  values are averages over (imagined or real) ensembles of random
  realizations. The conceptual weakness of} {\color{red} Huygens's
  Criterion} {\it \color{red}is its limited relevance to an}
          {\color{red} individual} {\it \color{red} making a
            decision. Either the individual has to be part of a large
            resource-sharing group mimicking a statistical
            ensemble\footnotemark, or the wealth process $W(t)$ has to
            be additive for the rate of change to be ergodic so that
            the expectation value reflects how the individual will
            fare over time. Wealth is often better modeled with
            multiplicative dynamics.}
\end{framed}
\footnotetext{The necessary size of this
              group grows exponentially in time for multiplicative
              dynamics \cite{Redner1990,PetersKlein2013}}

Specifically, N. Bernoulli proposed the following lottery: a fair coin
is tossed until heads appears for the first time. The number of coin
tosses, $n\in \{1, 2...\}$, is modeled as a random variable with
geometric distribution, $p_n(n)=2^{-n}$. The payout as a function of
$n$ is $D(n)=\$2^{n-1}$.
It follows that $D(n)$ is power-law distributed with diverging first
moment.
The time $\Delta t$ to generate an instance of
the random variable, \ie to play the lottery, is considered
independent of $n$ in this study. The lottery is usually presented
without restriction on $n$. For a more careful treatment of the problem one
must initially require $n\leq\nmax$. For more than $\nmax$ coin tosses the lottery
is declared invalid and no change in wealth occurs. The behavior of the 
original unrestricted case is investigated as the limit $\nmax\to\infty$.

\subsection{1738--1814 decision theory -- utility}
\label{1738}

By 1738 N. Bernoulli's cousin Daniel Bernoulli and Cramer
\cite[p.~33]{Bernoulli1738} had conceptualized the problem as
follows. They argued that people attach a value to money that is
non-linear in the dollar amount. Cramer had written to N. Bernoulli in
1728: ``in their theory [\ie {\it Huygens' Criterion}] mathematicians
evaluate money in proportion to its quantity while, {\it in practice},
people with common sense evaluate money in proportion to the utility
they can obtain from it.''  Bernoulli suggested a logarithm to map a
dollar amount into utility \footnote{We know today that this conceptualization is
  problematic. From the requirement that there be no distinguished
  currency follows that a change in currency by some factor must lead
  to a corresponding change in any physical quantity that is some
  power of that factor, but the logarithm is not a power law, see
  \cite[p.~17 ff]{Barenblatt2003}.}
 $U_B(W)=\ln(W)$.
The quantity, Bernoulli suggested,
that people consider when deciding whether to take part in the lottery
is a combination of the expected gain in their utility if no ticket
price were paid, and the loss in utility they suffer when they pay the
ticket price. This leads to

\underline{\it \bf Bernoulli's Criterion:} \\
{\it a lottery ticket is worth buying if 
the following quantity is positive \cite[pp.~26--27]{Bernoulli1738}:
\begin{equation}
\elabel{Bernoulli_criterion}
\ave{\Delta U_B^+} - \Delta U_{B}^{-}=\sum_{n}^{\nmax}p_n(n)
\ln\left(\frac{W+D(n)}{W}\right) - \ln\left(\frac{W}{W-P}\right).
\end{equation}}
The first terms on either side of the equation represent the expected
gain in logarithmic utility, resulting from the payouts of the
lottery. This would represent the net change in utility if tickets
were given away for free, $P=0$. The second terms represent the loss
in logarithmic utility suffered at the time of purchase, \ie after the
ticket is bought but before any payout from the lottery is received.
This is inconsistent with expected-utility theory, as was pointed out
in \cite{Peters2011b}. 
The conceptual inconsistency may be phrased as follows: Bernoulli
thought of utility as a new currency, but currency conversion is
always linear -- at odds with the non-linear logarithm favored by
Bernoulli.
\begin{framed}
\underline{\it \bf Bernoulli's inconsistency}\\ {
  \color{red}Bernoulli's Criterion }{\it \color{red} is mathematically
  inconsistent with later work in expected-utility theory because
  Bernoulli did not calculate the expected net change in logarithmic
  utility. He did not only replace money with utility of money but
  also computed an observable other than the expected change in this
  new object.}
\end{framed}

\subsection{1814--1934 decision theory -- expected utility}

The consensus in the literature on utility theory is that Bernoulli
meant to compute the expected net change in utility and made a slight
error. Laplace \cite{Laplace1814} re-told Bernoulli's resolution of the St
Petersburg paradox and the invention of utility. Perceiving {\it
  Bernoulli's Criterion} as an error, he implicitly ``corrected''
Bernoulli's formal inconsistency without mention.

\underline{\it \bf Laplace's Criterion:} \\ {\it Maximize the expected rate
of change in logarithmic utility \cite[pp.~439--442]{Laplace1814},}
\begin{equation}
\elabel{Laplace_criterion}
\frac{1}{\Delta t}\ave{\Delta U_B(W)}=\frac{1}{\Delta t}\sum_{n}^{\nmax}p_n(n)
\left[\ln(W+D(n)-P)-\ln(W)\right].
\end{equation}
Later researchers adopted Laplace's corrected
criterion. Todhunter \cite{Todhunter1865} followed Laplace, as do modern
textbooks in stating that utility is an
object encoding human preferences in its expectation value, \eg
\cite{vonNeumannMorgenstern1944,ChernoffMoses1959,Samuelson1983}.
Laplace stayed within Bernoulli's conceptual framework and was almost
certainly not aware of the physical interpretation of his criterion
as the ergodic growth rate under multiplicative dynamics
\eref{ergodic_mult}. 

Bernoulli motivated the logarithm by suggesting that the perceived
utility change induced by an extra dollar is inversely proportional to
total wealth, leading to the differential equation $dU(W)=1/W$, whose
solution is the logarithm. But Bernoulli also considered Cramer's
suggestion of $U_C=\sqrt{W}$ a good representation of diminishing
marginal utility. The modern perspective takes Bernoulli's logarithm
more seriously than he himself did. The route to the modern treatment
of the gamble problem is to ask: ``what if the logarithm was not
merely convenient and a good fit to the data, what would be its
physical meaning if it truly was a logarithm?''  Using the logarithm
in exactly the same place as the utility function is equivalent to
assuming multiplicative dynamics and constructing an ergodic
observable.

\subsection{Post-1934 decision theory -- bounded utility}
\label{Summary}
Karl Menger \cite{Menger1934} re-visited Bernoulli's 1738 study, and came to
the incorrect conclusion that only bounded utility functions are
permissible. Of course, whether a utility function, or anything else,
is bounded or not in the limit of diverging wealth is practically
irrelevant because financial wealth will always be represented by a
finite number, as was pointed out \eg by Coolidge \cite{Coolidge1925}. However,
based on formal arguments Menger drew conclusions for the structure of
the permissible formalism, namely he ruled out linear and logarithmic
functions as models of behavior, and, equivalently, additive and
multiplicative processes as models of wealth. Because of the central
role of these dynamical models the development of decision theory
suffered from this restriction, and it is satisfying to see that
formal arguments against these important models are invalid, as
intuition would suggest. Menger must have been unaware of the
correction to Bernoulli's work by Laplace. His error may be phrased as
using {\it Bernoulli's Criterion} instead of Laplace's, and only considering
the first term in {\it Bernoulli's Criterion}, implicitly setting the ticket
price to zero, $P=0$. The invalidity of Menger's claim was pointed out
in \cite{Peters2011b}, for a detailed discussion, see
\cite{Peters2011c}. Menger's argument survives as received
wisdom. For completeness, we state it here and specify the invalid
inferences involved.
\vfill\eject
\begin{framed}
\underline{\bf Menger's flawed argument}
\begin{enumerate}
\color{red}
\item Logarithmic utility resolves the original St Petersburg paradox
because it turns exponentially increasing wealth-payouts,
$D_n\propto \exp(n)$, into linearly increasing
utility-payouts $\Delta U(n)\propto n$ for large $n$.
\item
If payouts increase even faster, \eg as the exponential of an
exponential, $\exp(\exp(n))$, then expected utility changes will diverge positively as
$\nmax$ diverges, just as expected wealth changes diverge for
exponentially increasing payouts.
\item
In such games logarithmic utility predicts that the player will be
willing to pay any ticket price, just as linear utility does for
exponentially increasing payouts. In this sense logarithmic utility is
not qualitatively different from linear utility. For utility theory to
achieve the desired generality, utility functions must be bounded.
\end{enumerate}
\end{framed}

The argument sounds plausible. If the logarithm specifies the value
attached to money, like another currency, then there is no intuitive
reason why it should be qualitatively different from a linear
function. But the logarithm encoding multiplicative dynamics provides
us with additional intuition: multiplicative dynamics imply an
absorbing boundary. Unlike under additive dynamics it is impossible to
recover from bankruptcy, and this is a qualitative difference. 
In the coin-toss example in \fref{coin_toss} bankruptcy cannot occur, but in 
the general gamble under multiplicative dynamics, bankruptcy is possible
if at least one possible outcome, $n^*$, say, leads to the loss of one's entire wealth, 
$\Delta W(n^*)=-W(t_0)$, so that the corresponding growth factor is $r(n^*)=0$. 
The absorbing state $W=0$ can be 
reached but not escaped from.
Closer
inspection of Menger's argument reveals that the issue is indeed more
nuanced than he thought.

We separate out the first term, for the smallest payout, and write the
expected utility change as
\begin{widetext}
\begin{equation}
\elabel{M1}
\ave{\Delta U_B(W)}=p_n(1) \ln\left(\frac{W+D(1)-P}{W}\right) +\sum_{n=2}^{\nmax}p_n(n)
\ln\left(\frac{W+D(n)-P}{W}\right).
\end{equation}
\end{widetext}
This form motivates the following evaluation of the three steps in
Menger's argument
\begin{framed}
\begin{enumerate}
\item
Apart from turning exponential wealth changes into linear utility
changes, logarithmic utility also imposes a no-bankruptcy
condition. Bankruptcy becomes possible at $P=W+D(1)$. Reflecting this,
the limit $\lim_{P\to W+D(1)}\ave{\Delta U_B(W)}$ is negatively
divergent for any $\nmax$.
\item
If payouts increase as the exponential of an exponential then the
expected utility change is positively divergent in the limit
$\nmax\to\infty$ {\it only for ticket prices satisfying
  $P<W+D(1)$}. The double-limit $\lim_{P\to
  W+D(1)}\lim_{\nmax\to\infty}\frac{1}{\Delta t}\ave{\Delta U_B(W)}$
results in the indeterminate form ``$-\infty + \infty$.''  Note that
the positive divergence only happens in the unrealistic limit
$\nmax\to \infty$, whereas the negative divergence happens at finite
$P$. The negative divergence is physically meaningful in that it
reflects the impossibility to recover from bankruptcy under
multiplicative dynamics.
\item
In such games logarithmic utility does not predict that the player
will want to pay any finite ticket price. Instead, it predicts that
the player will not pay more than $W+D(1)$, irrespective of how $D(n)$
may diverge for large $n$. This is qualitatively different from
behavior predicted by {\it Huygens' criterion} (linear utility), where under
diverging expected payouts no ticket price exists that the player
would not be willing to pay. Logarithmic utility, carefully
interpreted, resolves the class of problems for which Menger thought
it would fail.
\end{enumerate}
\end{framed}

Despite a persisting intuitive discomfort, renowned economists
accepted Menger's conclusions and considered them an important
milestone in the development of utility theory. Menger implicitly
ruled out the all-important logarithmic function that connects utility
theory to information theory \cite{Kelly1956,CoverThomas1991} and
provides the most natural connection to the ergodicity argument we have presented.
Menger also ruled out
the linear function that corresponds to {\it Huygens' Criterion},
which utility theory was supposed to generalize.

Requiring boundedness for utility functions is methodologically
inapt. It is often stated that a diverging expected utility
is ``impossible'' \cite[p.~106]{ChernoffMoses1959}, or that it
``seems natural'' to require all expected utilities to be finite
\cite[p.~28--29]{Arrow1974}. Presumably, these statements reflect the
intuitive notion that no real thing can be infinitely useful. To
implement this notion in the formalism of decision theory, it was
decided to make utility functions bounded. A far more
natural way to implement the same notion would be to recognize that
money amounts (and quantities of anything physical, anything money
could represent) are themselves bounded, and that this makes any
usefulness one may assign to them finite, even if utility functions
are unbounded. There is no need to place bounds on $U(W)$ if $W$
itself is bounded.

\section{Summary and conclusion}
\label{Summary_and}
Our method starts by recognizing the inevitable non-ergodicity of
stochastic growth processes, \eg noisy multiplicative growth. The
specific stochastic process implies a set of meaningful 
observables with ergodic properties, \eg the exponential growth rate. These observables make
use of a mapping that in the tradition of economics is viewed as a
psychological utility function, \eg the logarithm.

The dynamic approach to the gamble problem makes sense of risk
aversion as optimal behavior for a given dynamic and and level of
wealth,
implying a different concept of rationality. Maximizing expectation
values of observables that do not have the 
ergodic property of Section~\ref{Preliminaries} cannot be considered
rational for an individual. Instead, it is more useful to consider rational the
optimization of time-average performance, or of expectation values of
appropriate ergodic observables. We note that where optimization is
used in science, the deep insight is finding the right object to
optimize (\eg the action in Hamiltonian mechanics, or the entropy in
the microcanonical ensemble). The same is true in the present case --
deep insight is gained by finding the right object to optimize -- we
suggest time-average growth.
{\it Laplace's Criterion} interpreted as an ergodic growth
rate under multiplicative dynamics avoids the fundamental circularity
of the behavioral interpretation. In the latter, preferences, \ie
choices an individual would make, have to be encoded in a utility
function, the utility function is passed through the formalism, and
the output is the same as the input: the choices an individual would
make.

We have repeated here that Bernoulli \cite{Bernoulli1738} did not actually
compute the expected net change in logarithmic utility, as was pointed
out in \cite{Peters2011b}. Perceiving this as an error,
Laplace \cite{Laplace1814} corrected him implicitly without mention. Later
researchers used Laplace's corrected criterion until Menger \cite{Menger1934}
unwittingly re-introduced Bernoulli's inconsistency and introduced a
new error by neglecting the second diverging term, $\Delta
U^-$. Throughout the twentieth century, Menger's incorrect conclusions
were accepted by prominent economists although they noticed, and
struggled with, detrimental consequences of the (undetected) error for
the developing formalism.

We have presented Menger's argument against unbounded utility
functions as it is commonly stated nowadays. This argument is neither
formally correct (it ignores the negative divergence of the
logarithm), nor compatible with physical intuition (it ignores the
absorbing boundary).  {\it Laplace's Criterion} -- contrary to common
belief -- elegantly resolves Menger-type games.

Logarithmic utility must not be banned formally because it is
mathematically equivalent to the modern method of defining an ergodic
observable for multiplicative dynamics. This point of view provides a
firm basis on which to erect a scientific formalism. The concepts we
have presented 
resolve the fundamental problem of decsision theory, therefore game
theory, and asset pricing. Cochrane's book \cite{Cochrane2001} is
important in this context as it sets out clearly that all of asset
pricing can be derived from the ``basic pricing equation'' --
precisely the combination of a utility function and expectation values
we have critiqued here. He further argues that the methods used in
asset pricing summarize much of macroeconomics. The problems listed
there as those of greatest importance to the discipline at the moment
can be addressed using the modern dynamic perspective.

In presenting our results we have made a judgement call between
clarity and generality. We have chosen the most general problem of
decision theory, but have treated it specifically for discrete time
and wealth changes. Gambles that are continuous in time and wealth
changes can be treated along the lines of \cite{Peters2011a}. The
specific St Petersburg problem was treated in detail in
\cite{Peters2011b}. We have contrasted purely additive dynamics with
purely multiplicative dynamics. A generalization beyond purely
additive or multiplicative dynamics is possible, just as it is
possible to define utility functions other than the linear or
logarithmic function. This will be the subject of a future
publication.  
The arguments we have outlined are not restricted
to monetary wealth but apply to anything that is well modeled by a
stochastic growth process. Applications to ecology and biology seem
natural.

\section*{Acknowledgments}
We thank K. Arrow for discussions that started at the workshop
``Combining Information Theory and Game Theory'' in 2012 at the Santa
Fe Institute, and for numerous helpful comments during the preparation
of the manuscript. OP would like to thank A. Adamou for discussions
and a careful reading of the manuscript.


\begin{table}
\caption{List of symbols}
\label{games}       
\begin{tabular}{ll}
\hline\noalign{\smallskip}
Symbol & Name and interpretation  \\
\noalign{\smallskip}\hline\noalign{\smallskip}
$t$ & time\\
$t_0$ & time before the first round of a gamble\\
$\Delta t$ & duration of one round of a gamble\\
$T$ & total number of sequential rounds of a gamble\\
$\tau$ & index specifying one sequential round of a gamble\\
$N$ & total number of parallel realizations of a gamble\\
$\nu$ & index specifying one parallel realization of a gamble\\
$n$  & integer specifying an outcome \\
$n^*$  & integer specifying an outcome that leads to bankruptcy\\
$\nmax$  & number of possible outcomes \\
$n_\tau$ & random outcome that occurs in round $\tau$\\
$W$  & wealth \\
$\overline{W}_{T}$  & finite-time average wealth \\
$W_\nu$& wealth in realization $\nu$\\
$\Delta W(n)$ & change in wealth from $t$ to $t+\Delta t$ if outcome $n$ occurs \\
$r(n)$ & growth factor if outcome $n$ occurs, $r(n)=\frac{W(t_0)+\Delta W(n)}{W(t_0)}$\\
$p_n(n)$  & probability of outcome $n$ \\
$D(n)$ & payout resulting from a lottery if outcome $n$ occurs\\
$\Delta U(n)$& change in utility resulting from outcome $n$\\ 
$P$ & price for a ticket in a lottery\\
$U$ & utility function\\
$U_C$ & Cramer's square-root utility function\\
$U_B$ & Bernoulli's logarithmic utility function\\
$\ave{\Delta U_B^+}$ &expectation value of gains in logarithmic\\ 
& utility at zero ticket price\\
$\Delta U_B^-$ & loss in logarithmic utility when reducing $W$ by $P$\\
$\ave{\cdot}$ & expectation value of $\cdot$\\
\end{tabular}
\end{table}
\FloatBarrier

\bibliography{./bibliography}

\begin{thebibliography}{10}

\bibitem{Arrow1974}
K.~Arrow.
\newblock The use of unbounded utility functions in expected-utility
  maximization: Response.
\newblock {\em Q. J. Econ.}, 88(1):136--138, 1974.

\bibitem{Barenblatt2003}
G.~I. Barenblatt.
\newblock {\em Scaling}.
\newblock Cambridge University Press, 2003.

\bibitem{Bernoulli1738}
D.~Bernoulli.
\newblock Specimen {T}heoriae {N}ovae de {M}ensura {S}ortis. {T}ranslation
  ``{E}xposition of a new theory on the measurement of risk'' by {L.} {S}ommer
  (1954).
\newblock {\em Econometrica}, 22(1):23--36, 1738.

\bibitem{Buchanan2013}
M.~Buchanan.
\newblock Gamble with time.
\newblock {\em Nature Phys.}, 9:3, January 2013.

\bibitem{ChernoffMoses1959}
H.~Chernoff and L.~E. Moses.
\newblock {\em Elementary Decision Theory}.
\newblock John Wiley \& Sons, 1959.

\bibitem{Cochrane2001}
J.~H. Cochrane.
\newblock {\em Asset Pricing}.
\newblock Princeton University Press, 2001.

\bibitem{Cohen1997}
E.~G.~D. Cohen.
\newblock Boltzmann and statistical mechanics.
\newblock In {\em {B}oltzmann's Legacy 150 Years After His Birth, {A}tti dei
  {C}onvegni {L}incei}, volume 131, pages 9--23. {A}ccademia {N}azionale dei
  {L}incei, Rome, 1997.

\bibitem{Coolidge1925}
J.~L. Coolidge.
\newblock {\em An introduction to mathematical probability}.
\newblock Oxford University Press, 1925.

\bibitem{CoverThomas1991}
T.~M. Cover and J.~A. Thomas.
\newblock {\em Elements of {I}nformation {T}heory}.
\newblock John Wiley \& Sons, 1991.

\bibitem{FermatPascal1654}
P.~Fermat and B.~Pascal.
\newblock Private correspondence between {Fermat} and {Pascal}, 1654.

\bibitem{Gell-MannHartle2012}
M.~Gell-Mann and J.~B. Hartle.
\newblock Decoherent histories quantum mechanics with one real fine-grained
  history.
\newblock {\em Phys. Rev. A}, 85:062120(1--12), 2012.

\bibitem{Huygens1657}
C.~Huygens.
\newblock {\em De ratiociniis in ludo aleae. ({O}n reckoning at Games of
  Chance)}.
\newblock T. Woodward, London, 1657.

\bibitem{Kelly1956}
J.~L. {Kelly~Jr.}
\newblock A new interpretation of information rate.
\newblock {\em Bell Sys. Tech. J.}, 35(4):917--926, July 1956.

\bibitem{Laplace1814}
P.~S. {Laplace}.
\newblock {\em Th\'{e}orie analytique des probabilit\'{e}s}.
\newblock Paris, Ve. Courcier, 2 edition, 1814.

\bibitem{Menger1934}
K.~Menger.
\newblock Das {U}nsicherheitsmoment in der {W}ertlehre.
\newblock {\em J. Econ.}, 5(4):459--485, 1934.

\bibitem{Montmort1713}
P.~R. Montmort.
\newblock {\em Essay d'analyse sur les jeux de hazard}.
\newblock Jacque Quillau, Paris. Reprinted by the American Mathematical
  Society, 2006, 2 edition, 1713.

\bibitem{Peters2011c}
O.~Peters.
\newblock Menger 1934 revisited.
\newblock {\em http://arxiv.org/abs/1110.1578}, 2011.

\bibitem{Peters2011a}
O.~Peters.
\newblock Optimal leverage from non-ergodicity.
\newblock {\em Quant. Fin.}, 11(11):1593--1602, November 2011.

\bibitem{Peters2011b}
O.~Peters.
\newblock The time resolution of the {St} {Petersburg} paradox.
\newblock {\em Phil. Trans. R. Soc. A}, 369(1956):4913--4931, December 2011.

\bibitem{PetersAdamou2013}
O.~Peters and A.~Adamou.
\newblock Stochastic market efficiency.
\newblock {\em SFI working paper 13-06-022}, 2013.

\bibitem{PetersKlein2013}
O.~Peters and W.~Klein.
\newblock Ergodicity breaking in geometric {B}rownian motion.
\newblock {\em Phys. Rev. Lett.}, 110(10):100603, March 2013.

\bibitem{Redner1990}
S.~Redner.
\newblock Random multiplicative processes: An elementary tutorial.
\newblock {\em Am. J. Phys.}, 58(3):267--273, March 1990.

\bibitem{Samuelson1983}
P.~A. Samuelson.
\newblock {\em Foundations of economic analysis}.
\newblock Harvard University Press, enlarged edition, 1983.

\bibitem{Tao2012}
T.~Tao.
\newblock {\em Topics in random matrix theory}.
\newblock {A}merican {M}athematical {S}ociety, 2012.

\bibitem{Todhunter1865}
I.~Todhunter.
\newblock {\em A history of the mathematical theory of probability}.
\newblock Macmillan \& Co., 1865.

\bibitem{vonNeumannMorgenstern1944}
J.~{von Neumann} and O.~Morgenstern.
\newblock {\em Theory of games and economic behavior}.
\newblock Princeton University Press, 1944.

\end{thebibliography}
\bibliographystyle{abbrv}

\end{document}